\documentclass[aps,prb,floatfix,twocolumn,superscriptaddress]{revtex4-2}
\usepackage{graphicx}
\usepackage[english]{babel}
\usepackage{amsmath}
\usepackage{amssymb}
\usepackage{tensor}
\usepackage{bbold}
\usepackage{bm}

\newcommand{\vk}{\mathbf{k}}

\newcommand{\be}{\begin{eqnarray}}
\newcommand{\ee}{\end{eqnarray}}
\newcommand{\p}{\partial}

\newcommand{\dc}{c^{\dagger}}
\newcommand{\da}{a^{\dagger}}

\def\ket#1{|#1\rangle}
\def\bra#1{\langle #1 |}
\def\ep#1{\langle #1 \rangle}

\begin{document}

\title{Proxy ensemble geometric phase and proxy index of time-reversal invariant topological insulators at finite temperatures}

\author{Aixin Pi}
\affiliation{College of Physics, Sichuan University, Chengdu, Sichuan 610064, China}
\email{heyan$_$ctp@scu.edu.cn}

\author{Ye Zhang}
\affiliation{College of Physics, Sichuan University, Chengdu, Sichuan 610064, China}

\author{Yan He}
\affiliation{College of Physics, Sichuan University, Chengdu, Sichuan 610064, China}
\email{heyan$_$ctp@scu.edu.cn}

\author{Chih-Chun Chien}
\affiliation{Department of Physics, University of California, Merced, CA 95343, USA.}
\email{cchien5@ucmerced.edu}

\begin{abstract}
The ensemble geometric phase (EGP) has been proposed as a topological indicator for finite-temperatures systems. The ensemble Wilson loop, or the transfer matrix, contains the crucial information in the EGP construction. We propose a proxy index and a proxy EGP directly from the transfer matrix and apply them to time-reversal invariant topological insulators exemplified by the Bernevig-Hughes-Zhang (BHZ) and Kane-Mele (KM) models. The quantized proxy index and proxy EGP smoothly generalize the ground-state topological index to finite temperatures. For the BHZ model, a comparison with another topological indicator, the Uhlmann phase, shows different transition behavior with temperature. For the KM model, the EGP have been generalized to the time-reversal EGP previously, but the proxy EGP does not require any splitting of the contributions. The proxy index and proxy EGP thus offer an efficient means for characterizing finite-temperature topological properties.
\end{abstract}

\maketitle

\section{Introduction}
The concepts of topology have played an important role in condensed matter physics for understanding various phases of quantum materials beyond the scope of the Landau symmetry-breaking paradigm \cite{Zhang_TIRev,Kane_TIRev,Chiu2016}. For free fermions, the topology is usually protected by certain discrete symmetries. It has been argued that topological materials host robust edge modes stable against symmetry preserving perturbations \cite{Wu06}. The topological stability has triggered enthusiasm of the thriving field \cite{Bernevig_book,ShenTI,Asboth2016,Stanescu_book}. Most  achievements of topological matter so far have been obtained under the condition of zero temperature. However, in the real world, finite temperatures or environmental couplings are almost inevitable. This naturally leads to attempts to understand the topology of mixed states or open quantum systems~\cite{Anandan88,Uhlmann,Sjoqvist00,Rezakhani06,Huang14,Viyuela14,Diehl18,Asorey19}. 

The ground-state topology is usually characterized by quantized topological indices, many of which are based on the notion of geometric phases accumulated during an adiabatic evolution in the parameter space. The Berry phase from the Berry connection \cite{Berry,GPhase_book,Bernevig_book}, for example, lays the foundation for constructing ground-state topological indices. Inspired by the idea of the Berry connection, several different approaches has been proposed to extend the concept of geometric phases from pure states to mixed states \cite{Uhlmann,Sjoqvist00,Rezakhani06,Viyuela14,Diehl,Diehl18}. For example, the Uhlmann connection and Uhlmann phase have been a promising finite-temperature topological indicator~\cite{Uhlmann,Uhlmann1,Uhlmann2}. The Uhlmann connection depends on a specified parallel condition in the space of purified states of the density matrix \cite{Uhlmann,Uhlmann1,Hubner}. This method has been applied to several one- or two-dimensional topological \cite{Viyuela14,Viyuela2,HeChern18} and spin models \cite{Galindo21,HouPRA21}, many of which exhibit finite-temperature topological transitions separating topologically trivial and non-trivial phases. Recently, it has been employed to describe time-reversal invariant topological insulators with a $Z_2$ index at finite temperatures \cite{Zhang2021}.

Here, we focus on another recently-proposed topological indicator for mixed states, called the ensemble geometric phases (EGP) \cite{Diehl18,Wawer-1}. The EGP is a direct extension of the ground-state expectations of the polarization or the translation operator to a thermal average. The detailed analysis in Ref. \cite{Diehl18} shows that the EGP depends on a ``transfer matrix'', which is actually the Wilson loop of the non-Abelian Berry connection of all the bands with multiple insertions of the Boltzmann factors. Due to the products of the Boltzmann factors, the lowest energy state will dominate the contributions of the EGP. Therefore, the EGP of systems in the thermodynamic limit will approach the Berry phase of the lowest energy band. This suggests that the EGP is guaranteed to be quantized if one integrates it along a closed loop in the parameter space. In contrast, the Uhlmann phase may not have such a property. Another feature of the EGP is that it basically remains the same for all finite temperatures, meaning that a topological transition only takes places at infinite $T$. This again differs from the Uhlmann phase, where finite-temperature topological transitions may be found in many systems~\cite{Viyuela14,HouPRA21,Zhang2021}. The EGP has been applied to the Chern insulator~\cite{Wawer-1} as an example. 

Recently, a generalization called the time-reversal EGP has been proposed \cite{Wawer-2} in order to analyze the time-reversal invariant Kane-Mele (KM) model~\cite{Kane-Z2} at finite temperatures. The time-reversal EGP introduces a splitting of the contributions from different bands, similar to the procedure of evaluating the spin Chern numbers~\cite{Wawer-2}. The concept is more involved and the computation can be demanding. As an alternative, we will construct a proxy index and a proxy EGP based on the above mentioned transfer matrix that gives the same information of time-reversal invariant topological insulators without splitting the band contributions. We test the proxy indicators on the Bernevig-Hughes-Zhang (BHZ) model \cite{BHZ-1} and the KM model to show consistency with the ground-state and finite-temperature EGP results. The BHZ model captures the main features of the HgTe quantum well, in which the quantum spin Hall effect was first experimentally observed \cite{BHZ-2}. The KM model is another example of time-reversal invariant topological insulator originally proposed for graphene, but the small spin-orbit coupling in graphene makes its verification more challenging. Nevertheless, it may be possible to realize the KM model in engineered systems~\cite{Jotzu14}.

The reason why the proxy index and proxy EGP works is that the transfer matrix can be thought of as a finite-temperature counterpart of the $T=0$ Berry Wilson loop, which has been used as a $Z_2$ index for the BHZ model in the ground state \cite{Yu-Z2} and may be measured via interferometry~\cite{Grusdt14}. The proxy index is constructed from the phases of the eigenvalues of the ensemble Wilson loop and serves as an indicator of the $Z_2$ index at finite temperatures. 
Moreover, the proxy EGP exhibits quantized values for the time-reversal invariant BHZ and KM models and reflects the $Z_2$ index at finite temperatures without splitting the band contributions.

The rest of the paper is organized as follows. Section \ref{sec-T0} briefly reviews the ground-state topological properties of the time-reversal invariant BHZ and KM models via the Wilson loop.  Section \ref{sec-EGP} reviews the derivation of the EGP via the ensemble Wilson loop, from which the proxy EGP is introduced. Section \ref{sec-num} presents the proxy index and proxy EGP of the BHZ and KM models at finite temperatures, along with comparisons with the Uhlmann phase and the time-reversal EGP. Possible implications and measurements are also discussed. Section \ref{sec-conclu} concludes our study. Details of some calculations and an exactly solvable model are summarized in the Appendix.

\section{Summary of time-reversal invariant topological insulators at $T=0$}\label{sec-T0}
We begin by briefly reviewing the ground-state properties of two  prototypical time-reversal invariant topological insulators described by the BHZ and KM models.

\subsection{BHZ model}
\label{sec-BHZ-T0}
The Hamiltonian of the BHZ model is given by
\be
H=\left(
    \begin{array}{cc}
      H_0(\vk) &  H_1 \\
      H_1^{\dag} & H_0^*(-\vk)
    \end{array}
  \right).
\ee
The corresponding wave function is $\psi=(\psi_{1\uparrow},\psi_{2\uparrow},\psi_{1\downarrow},\psi_{2\downarrow})^T$, where the indices $i=1,2$ label the two orbitals and the arrows label the two spins. $H_0$ is the Qi-Wu-Zhang model \cite{Qi1}, given by
\be
H_0=\sin k_x \sigma_1+\sin k_y \sigma_2+(m+\cos k_x+\cos k_y)\sigma_3.
\label{QWZ}
\ee
Here $\sigma_i$ for $i=1,2,3$ are the Pauli matrices. The model given by $H_0$ is a Chern insulator with nonzero Chern number for $-2<m<2$. The $H_1$ term is given by
\be
H_1=\left(
    \begin{array}{cc}
      0 &  \gamma \\
      -\gamma & 0
    \end{array}
  \right),
\ee
which breaks the $S_z$ conservation and inversion symmetry. 

The topology of the BHZ model is protected by time-reversal symmetry, where the time-reversal (TR) operator is $U_T=i\sigma_2 K$ with $K$ denoting the complex conjugation. The TR invariant is given by
$U_T^\dag H^*(\vk) U_T=H(-\vk)$.
Due to TR symmetry, the lowest two bands are degenerate at the four time-reversal invariant momenta $\vk_1=(0,0)$, $\vk_2=(\pm\pi,0)$, $\vk_3=(0,\pm\pi)$ and $\vk_4=(\pm\pi,\pm\pi)$. The degeneracy fails the definition of the Chern number for each band, but the total Chern number of these two bands can still be defined. However, TR symmetry leads to a vanishing total Chern number. To reveal the non-trivial topology, one can introduce a $Z_2$ index. The Fu-Kane invariant \cite{Fu-Z2}
requires the use of globally defined eigenstates, which can be difficult to find in practice. As an alternative, we employ a manifestly gauge-invariant method based on the Wilson loop or Wannier center \cite{Yu-Z2}. In the continuum limit, the Wilson line can be expressed in terms of the non-Abelian Berry connection as
\be
&& W_{i,i+1}(k_y)\approx\exp\Big(i A_x(k_{x,i},k_y)\Delta k\Big), \\
&& A^{mn}_{\mu}(\vk)=-i\ep{u_m(\vk)|\frac{\p}{\p k_{\mu}}|u_n(\vk)}.  \nonumber
\ee
Here $\mu=x,y$ and $\Delta k=k_{x,i+1}-k_{x,i}$. The Wilson loop is then given by
\be
W(k_y)=\mathcal{P}\exp\Big(i\oint_C A_{\mu}(\vk)d k_{\mu}\Big).
\ee
Here $\mathcal{P}$ denote the path order of the following integral. The integral contour $C$ is the loop with fixed $k_y$, and $k_x$ varies from $0$ to $2\pi$. 

For numerical calculations, we discretize the momentum space into a lattice and define a Wilson line operator across a given link on the lattice, whose matrix element is given by
\be
W^{mn}_{i,i+1}(k_y)=\ep{u_m(k_{x,i},k_y)|u_n(k_{x,i+1},k_y)}.
\ee
Here $\ket{u_m}$ denotes the eigenstate in momentum space and the indices $m,n$ run through all the occupied bands. In the case of the half-filled BHZ model, $W_{i,i+1}$ is a 2 by 2 matrix. Then the Wilson loop with fixed $k_y$ is the product of a series of links, given by
\be
W(k_y)=W_{0,1}W_{1,2}W_{2,3}\cdots W_{N-1,N}W_{N,0}.
\label{W-k}
\ee
Here $N$ is the lattice number along the $x$-axis. The Wilson line operator, however, is not gauge invariant. Under the transformation $\ket{u(\vk)}\to\ket{u_n(\vk)}e^{i\theta(\vk)}$, we find that $W_{i,i+1}\to W_{i,i+1}e^{i\theta(k_{x,i+1},k_y)-i\theta(k_{x,i},k_y)}$. Nevertheless, for a closed loop in Eq. (\ref{W-k}), all the gauge dependence cancels out, and the Wilson loop is manifestly gauge invariant.

\begin{figure}
\centering
\includegraphics[width=\columnwidth]{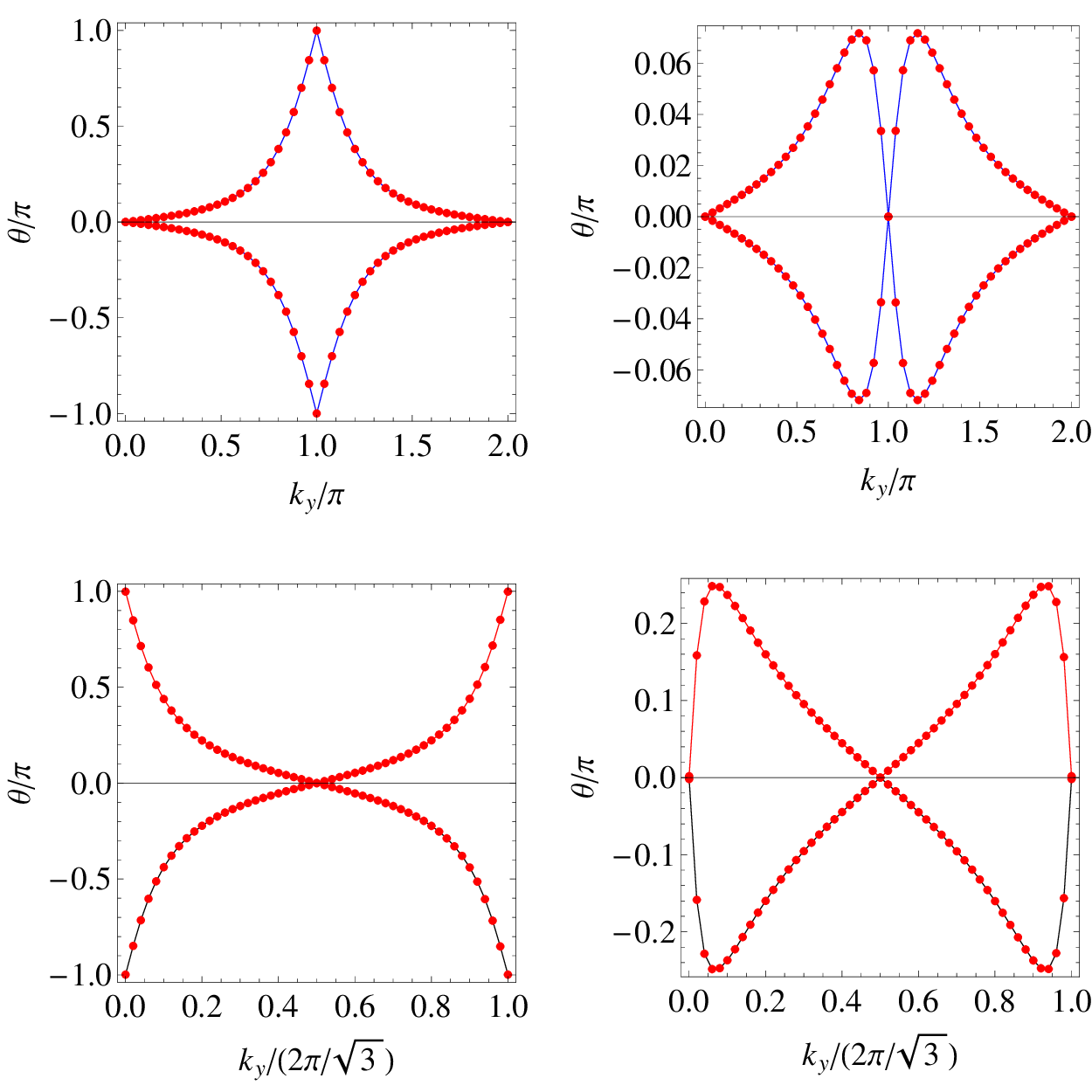}
\caption{The phases $\theta_1=-\theta_2$ of the eigenvalues of the Wilson loop $W(k_y)$ shown in Eq. (\ref{W-k}) for the BHZ model (top row) and the KM model (bottom row) as a function of $k_y$.  For the BHZ model, $\gamma=0.2$ and $m=1.5$ ($m=2.5$) for the left (right) panel. For the KM model, $\lambda_{SO}=0.06$ and $\lambda_v=0.1$ ($\lambda_v=0.4$) for the left (right) panel.}
\label{W-loop}
\end{figure}

With the Wilson loop, one can use the phases of the eigenvalues $\lambda_n$ of $W(k_y)$ to infer the $Z_2$ index as follows. We first obtain
\be
\theta_n(k_y)=\arg [\lambda_n(k_y)].
\ee
Here arg denote the phase angle or argument of a complex number.
Since $W(k_y)$ is a unitary matrix, its eigenvalues $\lambda_n(k_y)$ are all unit-modulus complex numbers.
For the BHZ model, there are only two arguments $\theta_{1,2}(k_y)$. Since det$(W)=1$, we always have $\theta_1=-\theta_2$. At the TR invariant momenta, $k_y=0$ or $k_y=\pi$, the Wilson loop $W$ has degenerate eigenvalues due to TR symmetry. Therefore, we have an additional condition $\theta_1=\theta_2$ at those points.

We plot the numerical result of the phases $\theta_{1,2}$ of the BHZ model in the top row of Figure \ref{W-loop}. In the left panel, we assume $m=1.5$ and $\gamma=0.2$ in the topological regime. Due to TR symmetry, $\theta_1=-\theta_2$ for all $\vk$. We have $\theta_{1,2}=0$ at $k_y=0$ and gradually increases to $\theta_{1,2}=\pm\pi$ at $k_y=\pi$. Note that $\pm\pi$ are the same modulo $2\pi$, so the requirement of $\theta_1=\theta_2$ at $k_y=\pi$ is satisfied. As $k_y$ further increases to $2\pi$, $\theta_{1,2}$ come back to zero. Therefore, $\theta_{1,2}$ wind around the whole $2\pi$, indicating the non-trivial topology. On the other hand, in the right panel with $m=2.5$ and $\gamma=0.2$ in the topologically trivial regime, we find that $\theta_{1,2}$ depart from zero not too far away before coming back to zero without any winding.

\subsection{KM model}\label{sec-KM-T0}
The KM model describes spin-1/2 fermions hopping on a honeycomb lattice with staggered sublattice potentials and spin-orbital couplings \cite{KM05}. Its Hamiltonian in momentum space can be written as
\be
&&H_{\text{KM}}=d_1\Gamma_1+d_2\Gamma_2+d_{12}\Gamma_{12}+d_{15}\Gamma_{15},\\
&&d_1=t(1+2\cos\frac{k_x}2\cos\frac{\sqrt{3}k_y}{2}), d_{12}=-2t\cos\frac{k_x}2\cos\frac{\sqrt{3}k_y}{2}, \nonumber\\
&&d_2=\lambda_v,
d_{15}=\lambda_{SO}(2\sin k_x-4\sin\frac{k_x}2\cos\frac{\sqrt{3}k_y}{2}). \nonumber
\ee
Here we have defined $\Gamma_1=\sigma_1\otimes s_0$, $\Gamma_2=\sigma_3\otimes s_0$, $\Gamma_{12}=-\sigma_2\otimes s_0$ and $\Gamma_{15}=\sigma_3\otimes s_3$. The matrices $\sigma_i$ and $s_i$ are the Pauli matrices in the sublattice space and spin space, respectively. The hopping coefficient $t$ can be taken as the energy unit. $\lambda_v$ and $\lambda_{SO}$ are the staggered potential strength and spin-orbital coupling (SOC), respectively. For a more clear comparison with the BHZ model, we start with the Rashba spin-orbital coupling set to zero and later discuss the case with finite Rashba terms.

The time-reversal (TR) operator for the KM model is $U_T K$ with $U_T=i\sigma_0\otimes s_2$ and $K$ denoting the complex conjugation. The KM model is time-reversal invariant since
$U_T^\dag H_{KM}^*(\vk) U_T=H(-\vk)$.
The KM model is topological when $\lambda_v<3\sqrt{3}\lambda_{SO}$ and becomes trivial if $\lambda_v>3\sqrt{3}\lambda_{SO}$. 
The phases of the eigenvalues of the Wilson loop of the KM model are shown in the bottom row of Fig.~\ref{W-loop}. Again, we have $\theta_1=-\theta_2$. Moreover, $\theta_1=\theta_2$ at the TR invariant momenta. Similar to the BHZ case, the phases wind around $2\pi$ in the topological regime and stay close to zero in the topologically trivial regime.

\section{Ensemble Geometric phase and its proxy}
\label{sec-EGP}
\subsection{Review of EGP}
The EGP $\varphi_E$ is defined as the polarization or the thermal average of many-body position operator $\hat{X}=\sum_j\hat{x}_j$ \cite{Diehl18}. However, the operator $\hat{X}$ is not a convenient choice for spatially periodic systems since it does not respect periodic boundary condition. It was pointed out by Resta  \cite{Resta98} that it is better to consider the translational operator $\hat{T}=\exp(i\delta k \hat{X})$, where $\delta k=2\pi/L$ with $L$ denoting the system size. Then, the polarization is just the phase of the thermal average of $\hat{T}$,
\begin{equation}
\varphi_E=\arg\langle\hat{T}\rangle \label{EGP}.
\end{equation}
Here $\langle \cdots\rangle=\textrm{Tr}(\rho \cdots)$ is the thermal average and $\rho$ is the density matrix. 

To compute the EGP defined in Eq. (\ref{EGP}), we consider a general tight-binding Hamiltonian in real space. In the second quantization form, the density operator can be expressed as
\begin{equation}
\rho=\frac{1}{\mathcal{Z}}\exp\Big(-\beta\sum_{i,j}\dc_i H_{ij} c_j\Big).
\end{equation}
Here $\dc_i$ and $c_i$ are the fermion creation and annihilation operators with $i$ collectively labels the lattice sites and the orbitals on each site. We assume that there are $N$ lattice sites and $n$ orbitals in total. The Hermitian matrix $H$ with elements $H_{i,j}$ is the Hamiltonian in the first quantized form and $\beta=1/k_B T$ with temperature $T$. We set $\hbar=1$ and $k_B=1$ throughout the paper. The normalization constant $\mathcal{Z}$ is included to ensure $\textrm{Tr}(\rho)=1$. Explicitly,
\begin{equation}
    \mathcal{Z}=\textrm{Tr}[\exp\Big(-\beta\sum_{i,j}\hat{c}_i^\dagger H_{ij}\hat{c}_j\Big)]=\det(\mathbb{1}+e^{-\beta H}).
\end{equation}
Meanwhile, in the second quantized form, the translation operator $\hat{T}$ becomes
\begin{equation}
\hat{T}(\dc,c)=e^{i\delta{k}\hat{X}}=e^{i\delta{k}\sum_i\dc_i x_i c_i}.
\end{equation}
In Ref. \cite{Diehl18}, the average $\ep{\hat{T}}$ is computed by using path integral. Instead, we make use of the following operator identity
\be
&&\textrm{Tr}\Big[\exp\Big(\sum_{ij}\dc_i X_{ij}c_j\Big)\exp\Big(\sum_{kl}\dc_k Y_{kl}c_l\Big)\Big]
=\det(\mathbb{1}+e^Xe^Y). \nonumber \\
&&
\label{eq-XY}
\ee
Here $\mathbb{1}$ is the identity matrix with the same dimension as $X$ and $Y$. The proof of this identity can be found in Appendix \ref{sec-XY}.
Applying the above formula to Tr$(\rho\hat{T})$, we find
\begin{equation}
   \begin{aligned}
\langle\hat{T}(\dc,c)\rangle
&=\frac{1}{\mathcal{Z}}\det(I_{nN}+\mathcal{T}\,e^{-\beta H}),
\label{EGP1}
\end{aligned}
\end{equation}
where we have defined a diagonal matrix $\mathcal{T}\equiv\text{diag}(e^{i\delta kx_1},\cdots, e^{i\delta kx_{N}})\otimes I_n$, and $I_n$ is the $n\times n$ identity matrix.

The real-space Hamiltonian can be block-diagonalized by transforming to momentum space as
\begin{equation}
H=\sum_k H_k\ket{k}\bra{k},
\end{equation}
where $H_k$ is an $n\times n$ Hermitian matrix defined in lattice momentum space. We can further diagonalize $H_k$ to find the eigen-energies as follows.
\begin{equation}
H_k=U_k E_kU_k^\dagger,\qquad E_k=\text{diag}(E_{1,k},\cdots,E_{n,k}),
\end{equation}
where $U_k$ is a unitary matrix whose columns are the eigenvectors of $H_k$. In momentum space, the matrix $\mathcal{T}$ can be expressed as
\begin{equation}
\mathcal{T}=\sum_k\ket{k+1}\bra{k}\label{egp2}.
\end{equation}
One can see that $\mathcal{T}$ only contains non-zero matrix elements on the upper sub-diagonal line.

Since $\mathcal{Z}$ is a real number and does not contribute to the phase, the EGP can be expressed as follows.
\begin{equation}
\begin{aligned}
\varphi_E&=\arg\det(\mathbb{1}+U e^{-\beta E}U^\dagger \mathcal{T})\\
&=\arg\Big[\exp\textrm{Tr}\Big(\sum_{n=1}^\infty (-1)^{n-1}\frac{A^n}{n}\Big)\Big].
\label{egp4}
\end{aligned}
\end{equation}
Here we have used the matrix identity $\det M=\exp(\textrm{Tr}\ln M)$ and also define $A=e^{-\beta E}U^\dagger \mathcal{T}U$ for convenience. In momentum space, $A=\sum A_k\ket{k-1}\bra{k}$, where $k=1,\ldots,N$. Some details of the evaluation can be found in Appendix~\ref{sec-EGPApp}.

For convenience, we introduce a path-ordered matrix product $M_T$, known as the transfer matrix \cite{Diehl18}, which is defined as
\be
&&M_T=(-1)^{N-1}\prod_k A_k=(-1)^{N-1}\prod_ke^{-\beta E_{k+1}}U_{k+1}^\dagger U_k. \nonumber \\
&&
\label{eq-Mt}
\ee
Here $U_{k+1}^\dagger U_k$ is the overlap between the eigenstates located at two adjacent points in momentum space. Collecting all the above results, we finally arrive at the following expression:
\be
\varphi_E&=&\arg\Big[e^{\textrm{tr}\ln(1+M_T)}\Big]=\arg\Big[\det(1+M_T)\Big].
\ee
This EGP expression was first derived in Ref. \cite{Diehl18}, but our derivation is slightly different. 

It is possible to simplify the expression of the transfer matrix $M_T$ by rewriting it as a path-ordered product of density matrices. Explicitly,
\be
M_T&=&\prod_i e^{-\beta E_{k_i}}U_{k_i}U_{k_i}^\dagger
=\prod_i \sum_n e^{-\beta E_{n,k_i}}\ket{u_n(k_i)}\bra{u_n(k_i)} \nonumber \\
&=&\prod_i \rho(k_i)
\ee
with unnormalized density matrix $\rho(k_i)$ at momentum $k_i$. In the continuum limit, it can also be written as a path-ordered integral as
\be
M_T=\mathcal{P}\exp\Big(-\frac{1}{T\Delta k}\oint_C H(\vk) d k_x\Big).
\label{MT-1}
\ee
Here $\Delta k$ is the difference between two adjacent momentum points. In order to find a meaningful $M_T$, we have to keep $\Delta k$ nonzero. Otherwise, the eigenvalues of $M_T$ will be either zero or infinite.

\subsection{Proxy ensemble geometric phases}
Although the EGP serves as a finite-temperature topological indicator related to the polarization, here we show that it is more convenient to use the transfer matrix $M_T$ as the central quantity for characterizing the topology. 
Inspired by the Uhlmann phase, which was previously used to characterize the topology of the two-dimensional Chern insulator at finite temperatures \cite{Viyuela2}, we propose
an alternative topological indicator called the proxy EGP, defined as follows.
\be
\Phi^E(k_y)=\arg\textrm{tr}\Big[M_T(k_y)\Big].
\ee
The proxy EGP is more computationally manageable since it directly extracts the information from $M_T$.
For a simple two-band system, the proxy EGP reproduces the EGP as follows. The two eigenvalues of $M_T$ satisfy $|\lambda_1|\gg1\gg|\lambda_2|$ due to the infinitely many products of the Boltzmann factors. In this case, the proxy EGP agrees with the EGP because
\be
\varphi^E=\arg\det(1+M_T)\approx\arg\lambda_1\approx\arg\textrm{tr} M_T=\Phi^E.
\ee

However, for models with degenerate or almost degenerate bands, the situation is more complicated, and the proxy EGP will be generally different from the EGP. Take the BHZ model for example, we will find the four eigenvalues of its $M_T$ satisfy 
$
|\lambda_1|\approx|\lambda_2|\gg1,~|\lambda_3|\approx|\lambda_4|\ll1.
$
Due to TR symmetry, $\lambda_1$ is the complex conjugate of $\lambda_2$. Hence,
$
\det(1+M_T)\approx\lambda_1\lambda_2=|\lambda|^2,
$
which is a positive number. Therefore, the EGP in this case is always zero. The analysis shows that a direct application of the EGP to time-reversal invariant topological models only leads to trivial results. As pointed out in Ref.~\cite{Wawer-2} with the KM model as an example, a possible way out is to define separate EGPs for the spin up and spin down bands. The difference of the spin up and down EGPs gives the time-reversal EGP, which can then detect the topology of time-reversal invariant systems.

On the other hand, a similar analysis leads to
$
\textrm{tr}M_T\approx\lambda_1+\lambda_2=2|\lambda_1|\cos\theta_1
$.
Here $\theta_1$ is the phase of $\lambda_1$. Therefore, depending on the value of $\theta_1$, the proxy EGP can take two quantized values $0$ or $\pi$ according to the sign of $\textrm{tr}M_T$. The analysis thus shows that the proxy EGP can directly reflect the underlying topology without splitting the contributions. In the following section, we will show that the proxy EGP indeed indicates the finite-temperature topology of time-reversal invariant topological insulators, using the KM and BHZ models as concrete examples. As will be shown shortly, the quantized values of the proxy EGP will lead to abrupt jumps indicating the non-trivial topology from the $Z_2$ index.

To calibrate our analytic and numerical results, we have verified the accuracy of the numerical calculations by an exactly solvable model summarized in Appendix~\ref{sec-solvable}.

\section{Results of time-reversal invariant topological insulators}\label{sec-num}

\subsection{BHZ model}
We first apply the formalism of the proxy EGP to the BHZ model at finite temperatures by focusing on the path-ordered matrix product $M_T$ in Eq. (\ref{eq-Mt}), which can be thought as a generalization of the Wilson loop to finite temperatures. To see this, we note that the overlap matrix can be expressed as
\be
&&(U^{\dag}_{k_{i+1}}U_{k_i})_{mn}=\ep{u_m(k_{i+1})|u_n({k_i})}\approx1-i\int_{k_i}^{k_{i+1}} A_{\mu}d k_{\mu},\nonumber \\
&& \\
&&A^{ab}_{\mu}(\vk)=-i\ep{u_a(\vk)|\frac{\p}{\p k_{\mu}}|u_b(\vk)}.
\ee
Here the indices $a,b$ indicate all possible orbitals $1,\cdots,n$. Making use of it, $M_T$ can be expressed as
\be
&&M_T=\prod_ie^{-\beta E(k_i)}\exp\Big(-i\int_{k_i}^{k_{i+1}} A_{\mu}d k_{\mu}\Big)\\
&&E_k=\text{diag}(E_{1,k},\cdots,E_{n,k}),\nonumber
\ee
which is just a path-ordered integral of the non-Abelian Berry connection with multiple insertions of the diagonal matrices of the Boltzmann factors. Here we have assume $N$ to be an odd number in order to drop the factor $(-1)^{N-1}$. It is now clear that $M_T$ may be referred to as the ensemble Wilson loop. If we replace the diagonal matrices $e^{-\beta E(k_i)}$ with another matrix diag$(0,\cdots,0,1,\cdots,1)$, where the non-vanishing elements correspond to the occupied bands, then we recover the Wilson loop of the non-Abelian Berry connections of those occupied bands.

\begin{figure}
\centering
\includegraphics[width=\columnwidth]{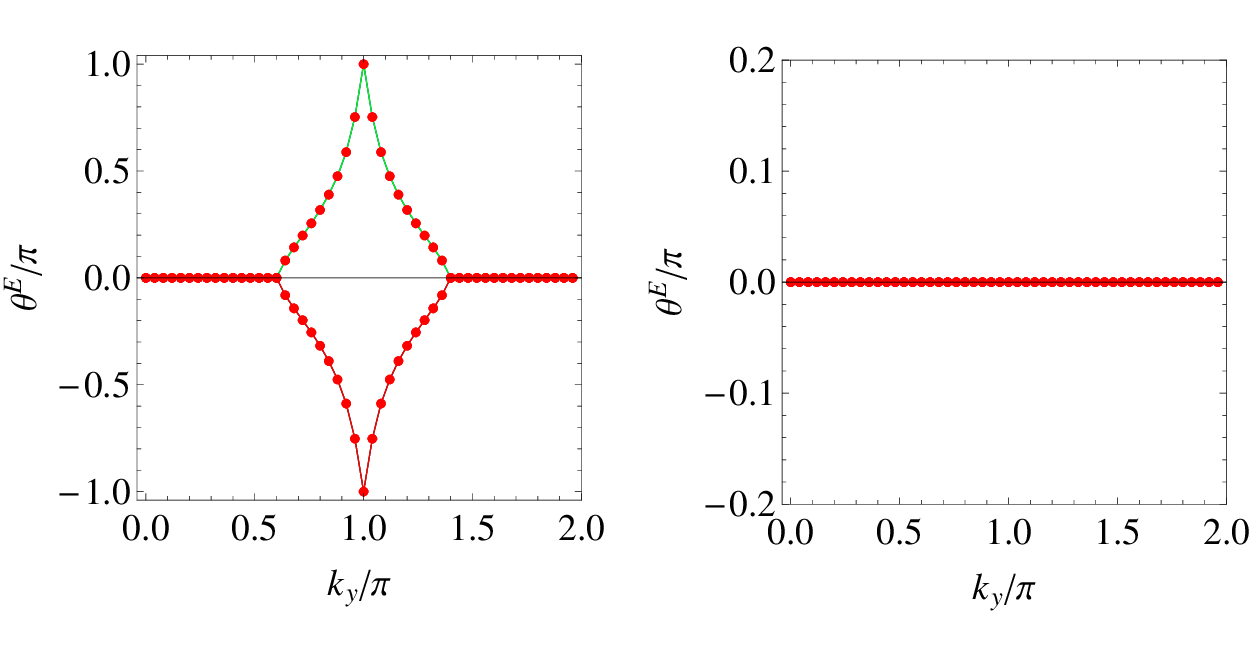}
\caption{The phase $\theta^E$ of the eigenvalues of $M_T(k_y)$ for the BHZ model as a function of $k_y$. Due to $\theta^E_1=\theta^E_2=-\theta^E_3=-\theta^E_4$, there are two sets of degenerate data on the plot. Here $\gamma=0.2$ and $T=5.0$ with $m=0.8$ and $m=2.8$ for the left and right panels, respectively.}
\label{EGP-th}
\end{figure}

$M_T$ is a 4 by 4 matrix for the BHZ model. At zero temperature, the topology is reflected by the phase of the eigenvalues of the Wilson loop. Analogously, we define the following phases for fixed $k_y$:
\be
\theta^E_n(k_y)=\arg \lambda_n(k_y).
\ee
Here $\lambda_n$ is the $n$-th eigenvalue of $M_T$. The numerical results of $\theta^E_n$ are shown in the top row of Figure \ref{EGP-th}. Note that the four $\theta^E$ satisfy the following relation
\be
\theta_1^E=\theta_2^E=-\theta_3^E=-\theta_4^E\label{th-E}.
\ee
Thus, although there are 4 eigenvalues, only two opposite phases are visible, making the plots similar to the Wilson-loop result from the Berry connection at $T=0$. In the left (right) panel of Figure \ref{EGP-th}, we assume $m=0.8$ ($m=2.8$) in the topological (trivial) regime. In both cases, a small spin-orbital coupling $\gamma=0.2$ and relative high temperature $T=5.0$ are assumed.

\begin{figure}
\centering
\includegraphics[width=\columnwidth]{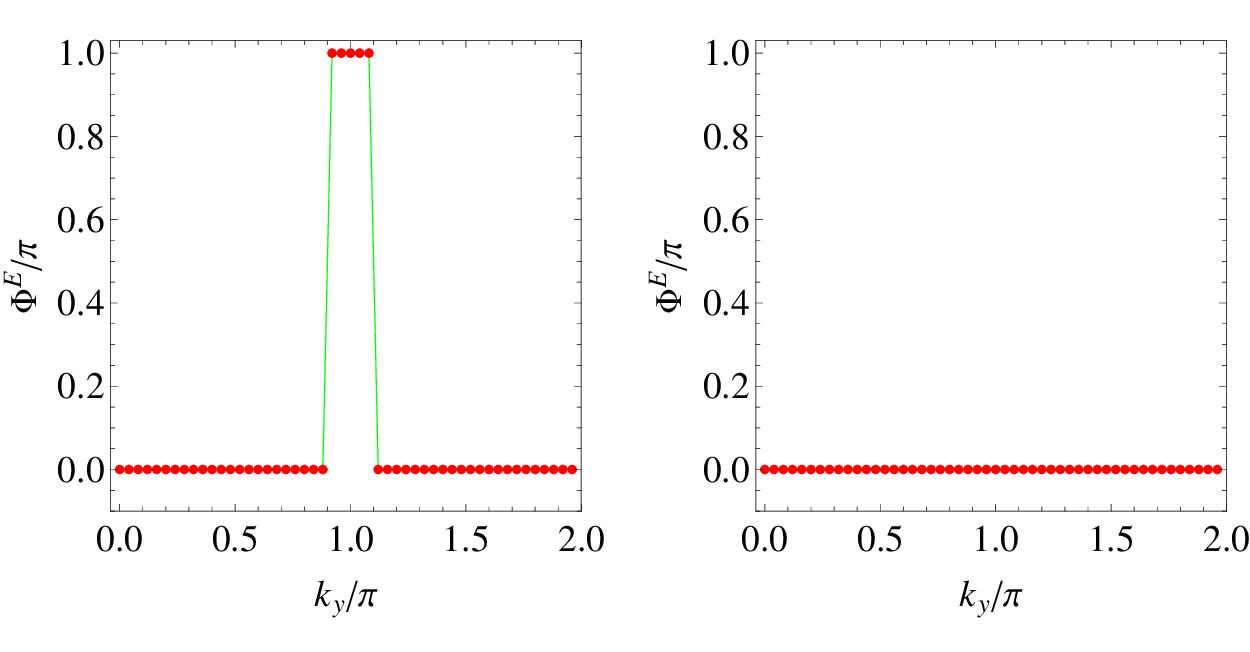}
\caption{The proxy EGP $\Phi^E$ for the BHZ model as a function of $k_y$. Here $\gamma=0.2$ and $T=5.0$ with $m=0.8$ (topological) and $m=2.8$ (trivial) for the left and right panels, respectively.}
\label{EGP-phi}
\end{figure}

A comparison of $\theta^E$ with $\theta$ from the Wilson loop at $T=0$ shows that they share similar qualitative features. They are both zero at $k_y=0$ or $k_y=2\pi$. In the topological regime, $\theta^E$ stays at zero and then starts to rise as $k_y$ increases. It reaches $\pm\pi$ at another TR invariant momentum $k_y=\pi$. Therefore, $\theta^E$ also winds around a full circle as we travel from $k_y=0$ to $k_y=\pi$. Thus, we can still interpret its winding as an indicator of the topology. When compared to Figure \ref{W-loop}, the only difference is that $\theta^E$ remains zero inside two small intervals of $k_y$ close to $0$ and $2\pi$. However, this does not affect the winding of the phase. On the other hand, for the topological trivial case, $\theta^E$ are basically zero. The comparisons show the phase of the eigenvalues of the ensemble Wilson loop can reflect the $T=0$ topological properties at finite temperatures. The winding of $\theta^E$ can also be inferred by the following index, which we call the proxy index for the EGP:
\be
\nu_n=\frac1{2\pi}\int_0^\pi d k_y\frac{\p}{\p k_y}\theta_n^E.
\ee
The topological regime corresponds to $\nu_n=\pm\frac12$ while the topologically trivial regime corresponds to $\nu_n=0$. The proxy index thus generalizes the $Z_2$ index to finite temperatures. We remark the definition is consistent with another $Z_2$ index called the magnetoelectric polarization defined at $T=0$ in Ref.~\cite{Qi08}.

We remark that we show the results for a selected value of $T=5$ as the qualitative behavior of $\theta^E$ are basically the same for all finite $T$. However, if we consider the limiting case of $T\to\infty$, $e^{-\beta E(k_i)}$ will be proportional to the identity matrix, causing $M_T$ to become the identity matrix as well. Therefore, $\theta^E_n=0$ for all $n$, and the system becomes topologically trivial at infinite high $T$ as expected. The topological transition according to the proxy index only occurs at $T=\infty$, which is different from the finite-temperature topological phase transition of the Uhlmann phase of the BHZ model~\cite{Zhang2021}.

The proxy EGP, $\Phi^E$, of the BHZ model as a function of $k_y$ is shown in Figure \ref{EGP-phi}. Similar to the Uhlmann phase of the BHZ model, $\Phi^E$ only takes two quantized values, $0$ or $\pi$. This is because the eigenvalues of $M_T$ of the BHZ model are all complex conjugate pairs, guaranteeing its trace to be a real number. Therefore, $\Phi^E$ is quantized at $0$ or $\pi$, forming a $Z_2$ group. For the topological case with $m=0.8$, $\Phi^E$ jumps from 0 to $\pi$ and then jumps back to zero later as $k_y$ goes from 0 to $2\pi$. On the other hand, $\Phi^E$ is always   the topologically trivial case with $m=2.8$. Therefore, the abrupt jump of $\Phi^E$ from $0$ to $\pi$ may serve as another indicator of finite-temperature topology. In both panels of Figure \ref{EGP-phi}, we assume $T=5.0$, but we have verified that the qualitative behavior of $\Phi^E$ remains the same for all finite $T$. Therefore, similar to the analysis of $\theta^E$, the topological transition only happens at infinite $T$.

When compared to the Uhlmann phase $\Phi^U$ of the BHZ model discussed in Ref.~\cite{Zhang2021}, the zero-temperature behavior of $\Phi^U$ and $\Phi^E$ is basically the same, reflecting the $Z_2$ index in the ground state. However, the regime where $\Phi^U=\pi$ can be found shrinks with $T$, and there is a finite-temperature topological phase transition to the trivial regime. In contrast, $\Phi^E$ stays qualitatively with its zero-temperature value for any finite temperature and only vanishes completely at infinite temperature. The two topological indicators thus give different pictures of finite-temperature topological properties of the BHZ model.

\begin{figure}
\centering
\includegraphics[width=\columnwidth]{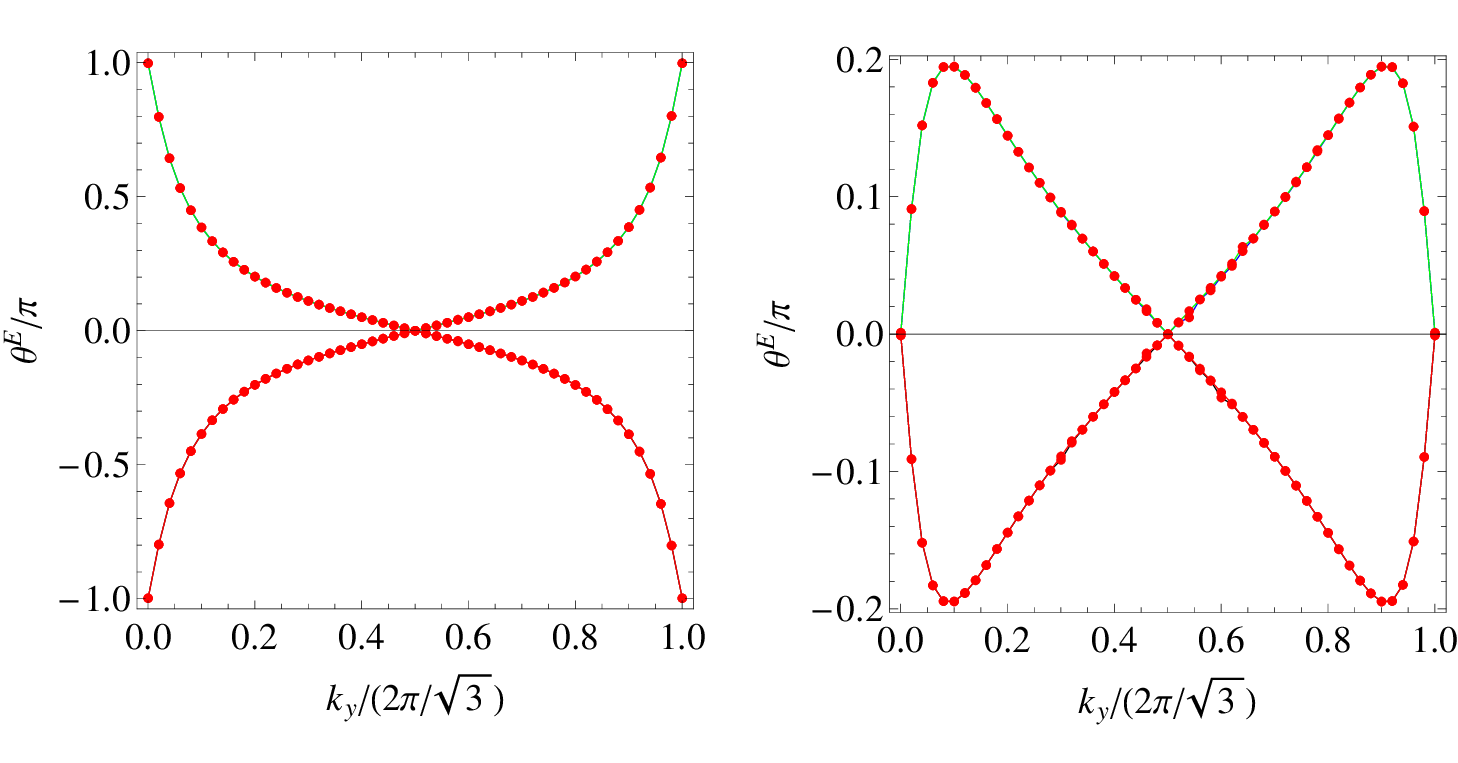}
\caption{The phases $\theta^E$ of the eigenvalues of $M_T(k_y)$ for the Kane-Mele model as a function of $k_y$. Due to $\theta^E_1=\theta^E_2=-\theta^E_3=-\theta^E_4$, there are two sets of degenerate data. Here $\lambda_{SO}=0.06t$ and $T=5.0t$ with $\lambda_v=0.1t$ (topological) and $\lambda_v=0.4t$ (trivial) for the left and right panels, respectively.}
\label{KM}
\end{figure}

\subsection{KM model}
The proxy index and proxy EGP from $M_T$ can also be applied to the Kane-Mele (KM) model. We plot the phase $\theta^E$ of the eigenvalues of $M_T(k_y)$ for the KM model as a function of $k_y$ in Figure \ref{KM}. Here $\theta_n^E$ also satisfy the relation of Eq.~(\ref{th-E}), so the data group into two degenerate sets of opposite values. In the left panel with small $\lambda_v$, $\theta^E$ winds around a full circle, representing the non-trivial topology. In the right panel with large $\lambda_v$, $\theta^E$ never finish a whole circle, showing the trivial case. Note that the range of $k_y$ is $(0,2\pi/\sqrt{3})$ due to the honeycomb lattice of the KM model. A proxy index of the winding number can also be defined for the KM model as
\be
\nu_n=\frac1{2\pi}\int_0^{\pi/\sqrt{3}} d k_y\frac{\p}{\p k_y}\theta_n^E.
\ee
Here the upper limit of the integral is different from that of the BHZ model because of lattice is not square. Again, $\nu_n=\pm1/2$ indicate the topological regime. Therefore, the proxy index generalizes the ground-state $Z_2$ index of time-reversal invariant topological insulators to finite temperatures.

\begin{figure}
\includegraphics[width=\columnwidth]{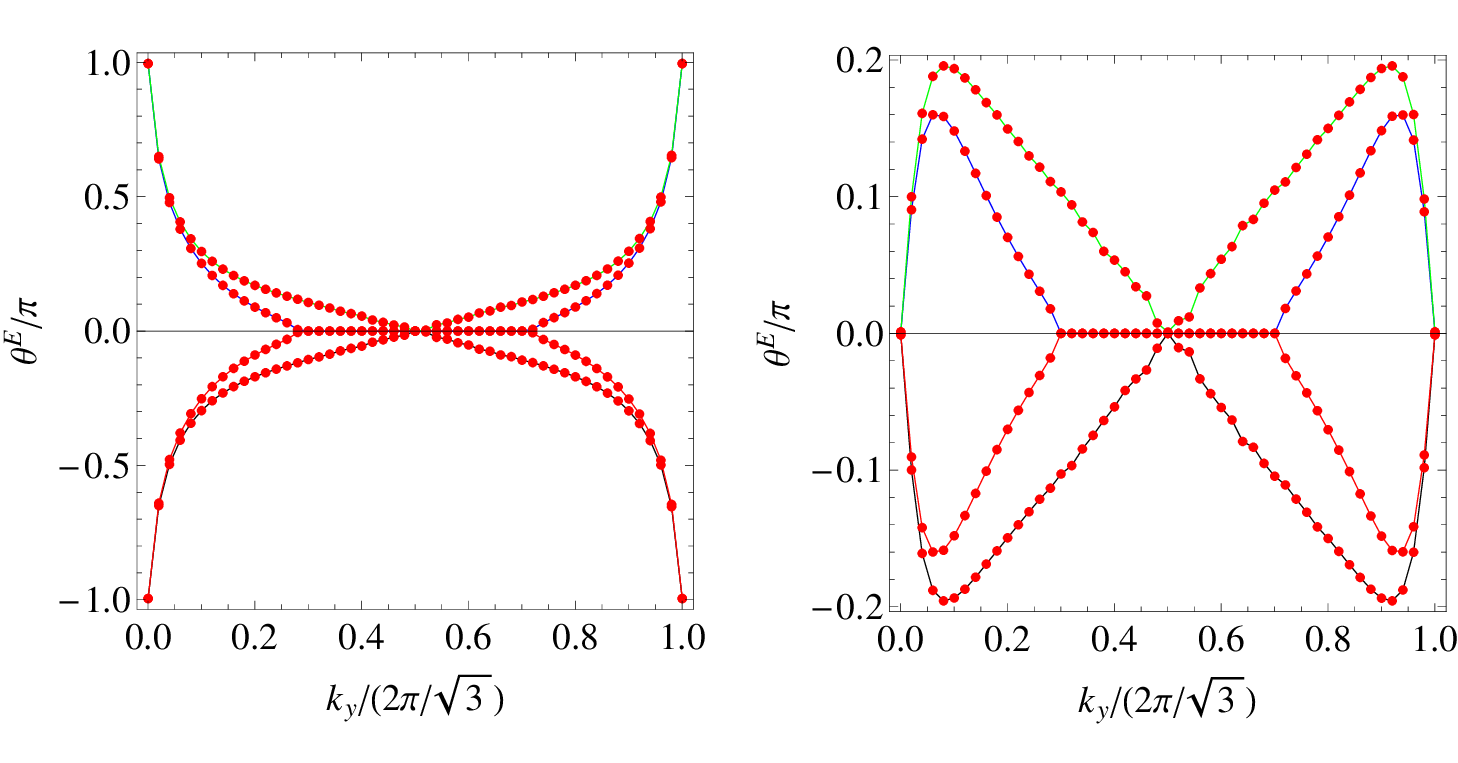}
\caption{The phases $\theta^E$ of the eigenvalues of $M_T(k_y)$ for the Kane-Mele model as a function of $k_y$. Here $\lambda_{SO}=0.06t$, $\lambda_R=0.1t$, and $T=5.0t$ with $\lambda_v=0.1t$ (topological) and $\lambda_v=0.3t$ (trivial) for the left and right panels, respectively.}
\label{KM-1}
\end{figure}

We also consider the KM model with the following Rashba type SOC term:
\be
&&H_{\text{Rashba}}=d_3\Gamma_3+d_4\Gamma_4+d_{23}\Gamma_3+d_{24}\Gamma_3,\\
&&d_3=\lambda_R(1-\cos x\cos y),
d_4=-\sqrt{3}\lambda_R\sin x\sin y, \nonumber\\
&&d_{23}=-\lambda_R\cos x\sin y,
d_{24}=\sqrt{3}\lambda_R\sin x\cos y. \nonumber
\ee
Here we have defined $x=k_x/2$, $y=\sqrt{3}k_y/2$, $\Gamma_3=\sigma_2\otimes s_1$, $\Gamma_4=\sigma_2\otimes s_2$, $\Gamma_{23}=-\sigma_1\otimes s_1$, and $\Gamma_{24}=\sigma_1\otimes s_2$.
The results of $\theta^E$ of the model are shown in Figure \ref{KM-1} for a moderate value of $\lambda_R=0.1t$. We find that with the finite Rashba coupling, the degeneracies between $\theta^E_j$s are lifted. Nevertheless, the qualitative behaviors are the same as the KM model without the Rashba term shown in Fig.~\ref{KM}. Thus, the proxy index is also the same.

\begin{figure}
\centering
\includegraphics[width=\columnwidth]{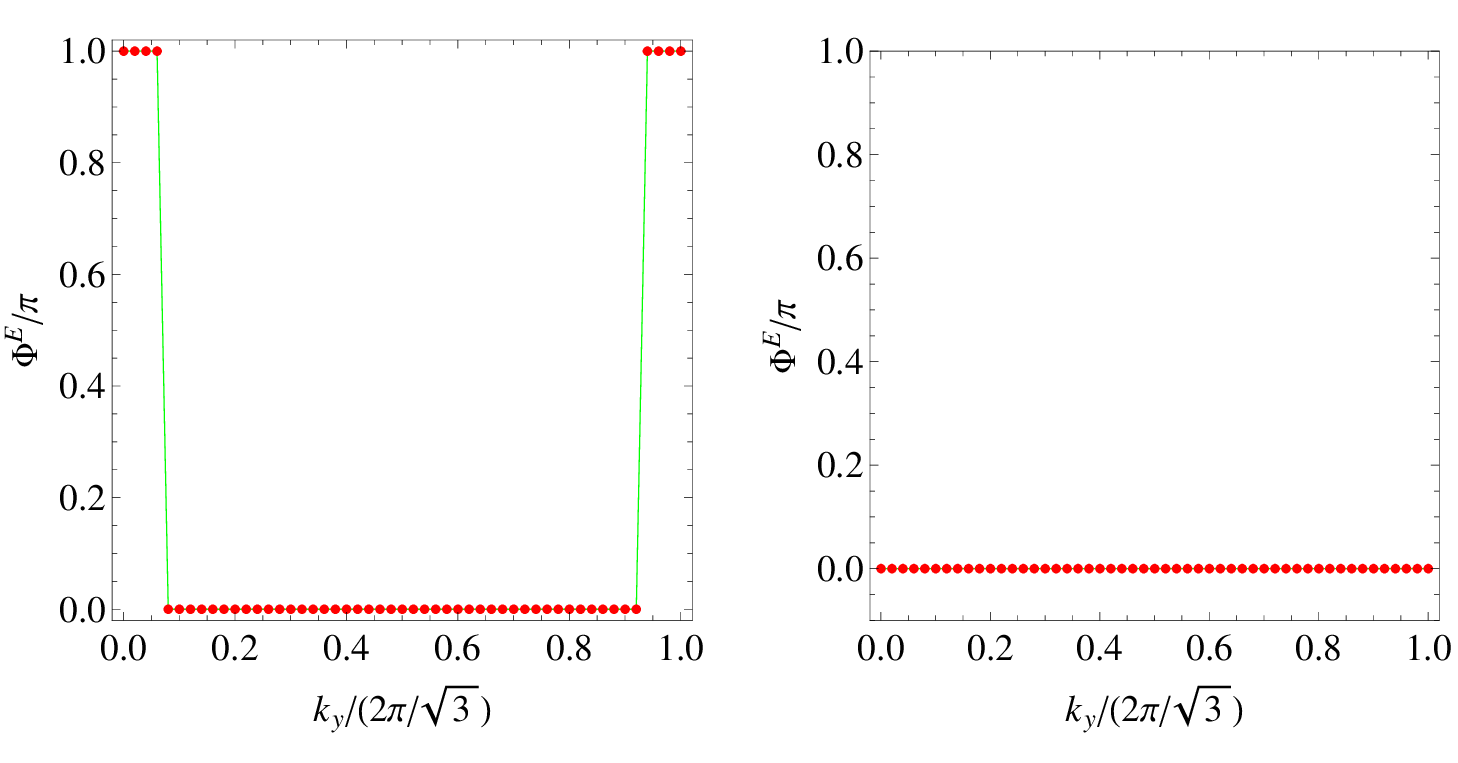}
\caption{The proxy EGP $\Phi^E$ for the Kane-Mele model as a function of $k_y$. Here $\lambda_{SO}=0.06t$ and $T=5.0t$ with $\lambda_v=0.1t$ and $\lambda_v=0.4t$ for the left and right panels, respectively.}
\label{KM-phi}
\end{figure}

Next, the proxy EGP, $\Phi^E$, of the KM model without the Rashba term is evaluated and shown in Figure \ref{KM-phi} as a function of $k_y$. Similar to the BHZ model, the eigenvalues of $M_T$ are all complex conjugate pairs, which also guarantee that the trace is a real number. Thus, $\Phi^E$ is quantized at $0$ or $\pi$. When $k_y$ is varied, finite $\Phi^E$ can be observed in the topological regime. The proxy EGP thus serves as another finite-temperature topological indicator, which does not require splitting of the band contributions as the time-reversal EGP discussed in Ref.~\cite{Wawer-2}. We also found that $\Phi^E$ remains quantized and behaves qualitatively the same when the Rashba SOC term with a moderate value is introduced.

\subsection{Implications}
The EGP is associated with the polarization, which may be measured experimentally from the expectation of the momentum shift operator~\cite{Wawer-1,Wawer-2}. Since the EGP smoothly extends the zero-temperature topological behavior to finite temperatures, there is no change of the topological regime as temperature increases, as long as the infinite temperature point is excluded. While the generalization to the time-reversal EGP allows a description of the finite-temperature behavior of the KM model~\cite{Wawer-2}, the splitting of band contributions may require additional care in the construction.

On the other hand, the proxy index and proxy EGP fully agree with the $T=0$ indicator and do not deviate from it at finite-temperatures when applied to the two models of time-reversal invariant topological insulators. Therefore, we do not consider the proxy index and proxy EGP as brand-new topological indicators in those cases. Rather, they provide a streamlined calculation to infer the finite-temperature topology characterized by the EGP in a computationally manageable manner. Experimentally, the EGP may be measured from the polarization or interferometry in natural or engineered materials~\cite{Wawer-1,Wawer-2} while the proxy index and proxy EGP allow for an efficient and unified characterization of the results.

We remark that different finite-temperature topological indicators reflect the robustness of different quantities against thermal averaging. Take the BHZ model as an example, the Uhlmann phase exhibits a finite-temperature transition to a trivial regime~\cite{Zhang2021}, showing the triviality of the holonomy in the Uhlmann bundle at high temperatures. On the other hand, the proxy index and EGP preserve the topological properties at any finite temperature through the Boltzmann factors. Both scenarios are valid, but they reflect different topological properties. We caution that for systems in two (or higher) dimensions, the calculation of the Uhlmann phase may show different results if the order of integrations are changed~\cite{Diehl}. As shown in Ref.~\cite{Zhang2021}, however, the BHZ model is a special case where the Uhlmann phase is insensitive to a change of the order of integrations. In contrast, the EGP and its proxies do not have such complications.

\subsection{Possible measurement of proxy EGP and proxy index}
Now we turn to possible ways for measuring the proxy EGP and proxy index. We first recall that the EGP is accessible by direct measurements through optical setups such as the Mach-Zehnder interferometer, as proposed in Ref.~\cite{Diehl18}. The idea of this measurement goes back to \cite{Sjoqvist00}.
We briefly review the main ideas of this type of EGP measurements. In a Mach-Zehnder interferometer, an incoming laser beam is split into two traveling along different paths, which may be called the upper lower arms. These photons in the two arms can be conveniently described by a two dimensional Hilbert space, which is the same as that of a qubit. In this language, the mirror reflection and  beam splitting can be described by 2 by 2 matrices as
\be
U_M=\left(
    \begin{array}{cc}
      0 & 1 \\
      1 & 0
    \end{array}
  \right),\quad
U_B=\frac{1}{\sqrt{2}}\left(
    \begin{array}{cc}
      1 & i \\
      i & 1
    \end{array}
  \right).
\ee

In the upper arm, the beam passes through a phase shift element, which delays the phase of the photon by $\chi$. Meanwhile, the beam in the lower arm passes through and interacts with a sample topological fermion system, which is the object that we want to probe. Now it is possible to engineer the interactions between the photons and fermions such that the photon beam in the lower arm will pick up a phase shift proportional to the center of mass of the fermions. This phase shift is just described by the argument of the expectation of the translation operator $\hat{T}=\exp(i\delta k \hat{X})$, which then gives the EGP. We refer the reader to Ref.~\cite{Diehl18} for the detailed construction of the interaction.
An the end, the two beams pass through a another beam splitter to give rise to the output signals. One can find the intensities of these two outputs, given by
\be
I_{\text{out}}^{\pm}=\frac12\Big(1\pm|\ep{\hat{T}}|\cos[\chi-\arg\ep{\hat{T}}]\Big)I_{\text{in}}.
\ee
By varying the phase $\chi$ and measuring the intensity change, one may deduce the phase $\arg\ep{\hat{T}}$, which is the EGP.

Now we come back to the proxy EGP and proxy index determined by the eigenvalues of $M_T$. The EGP only provides us information of $\varphi^E=\det(1+M_T)$, which is not enough to determine all the eigenvalues of $M_T$. In order to make use of the above EGP measurement, we propose an indirect way to determine the eigenvalues of $M_T$. We introduce a parameter to the momentum displacement by the replacement $\delta k\to \delta k/\eta$. According to Eq. (\ref{MT-1}), we find that $M_T$ will become $(M_T)^\eta$. By varying $\eta$, we can measure a series of EGP as
\be
(\varphi^E)_\eta=\det(1+M_T^\eta)=\prod_{j=1}^n(1+\lambda_j^\eta).
\label{exp}
\ee
Here $\lambda_j$ for $j=1,\cdots n$ are the eigenvalues of $M_T$. Now we can choose $n$ different values of $\eta$ to obtain $n$ different equations of the form of Eq. (\ref{exp}). Since $(\varphi^E)_\eta$ can be experimentally measured, we can in principle obtain all $\lambda_j$s from those $n$ equations. With all the $\lambda_j$s at hand, it is straightforward to determine the proxy EGP $\Phi^E=\arg(\sum_j\lambda_j)$ and the proxy index from $\theta^E_j=\arg(\lambda_j)$. For multi-band systems, however, the above procedure may be more challenging. We remark that the EGP has a more natural tie to physical quantities since it comes from the translation operator while the proxy EGP and index provide more straightforward evaluations of the time-reversal invariant topological systems discussed here.

\section{conclusion}
\label{sec-conclu}
Based on the EGP formalism, we propose the proxy index and proxy EGP via the ensemble Wilson loop or transfer matrix $M_T$ for characterizing time-reversal invariant topological insulators at finite temperatures, exemplified by the BHZ and KM models. The phases of the eigenvalues of $M_T$ display similar behavior as those of the Berry Wilson loop at $T=0$. The proxy index reflects the winding and distinguishes the topological and trivial phases. The proxy EGP is quantized for both BHZ and KM models and exhibits jumps between $0$ and $\pi$ in the topological regime. Moreover, the proxy EGP characterizes the time-reversal invariant topological insulators without the need for splitting the band contributions. The EGP and its proxies are free from complications of integration order that affects the Uhlmann phase in higher dimensions. Different from the Uhlmann phase of the BHZ model showing a finite-temperature topological transition, the proxy EGP only exhibits a transition at infinite temperature. Our study thus shows the rich physics of topological systems at finite temperatures.

\begin{acknowledgments}
Y. H. was supported by the Natural Science Foundation of China under Grant No. 11874272 and Science Specialty Program of Sichuan University under Grant No. 2020SCUNL210. C. C. C. was supported by the National Science Foundation under Grant No. PHY-2011360.
\end{acknowledgments}

\appendix
\section{Derivation of Eq. (\ref{eq-XY})}
\label{sec-XY}
Here we prove the identity of Eq. (\ref{eq-XY}). For convenience, we repeat this equation as follows.
\be
\textrm{Tr}\Big[\exp\Big(\sum_{ij}\dc_i X_{ij}c_j\Big)\exp\Big(\sum_{kl}\dc_k Y_{kl}c_l\Big)\Big]
=\det(\mathbb{1}+e^Xe^Y).\nonumber
\ee
We will first prove the following simple case:
\be
\textrm{Tr}\Big[\exp\Big(\sum_{ij}\dc_i X_{ij}c_j\Big)\Big]=\det(\mathbb{1}+e^X)
\label{eq-X}
\ee
Assume that the $N\times N$ matrix $X$ can be diagonalized as
\be
X=U^\dag \Lambda U,\quad \Lambda=\textrm{diag}(\lambda_1,\cdots,\lambda_N)
\ee
Here $\lambda_i$ with $i=1,\cdots,N$ are the eigenvalues of $X$. We can introduce a new set of fermion operators
\be
a_i=\sum_j U_{ij}c_j,\quad \da_i=\sum_j \dc_j U^\dag_{ji}
\ee
Then the right hand side of Eq.(\ref{eq-X}) can be simplified as
\be
&&\textrm{Tr}\Big[\exp\Big(\sum_{ij}\dc_i X_{ij}c_j\Big)\Big]
=\textrm{Tr}\Big[\exp\Big(\sum_{i}\lambda_{i}\da_i a_i\Big)\Big]\nonumber\\
&&=\prod_i\textrm{Tr}\Big[\exp\Big(\lambda_{i}\da_i a_i\Big)\Big]
=\det(\mathbb{1}+e^X).
\ee
This completes the proof of Eq. (\ref{eq-X}).

To prove Eq. (\ref{eq-XY}), we make use of the integral form of the Baker-Campbell-Hausdorff formula \cite{Hall-Book} as follows.
\be
Z\equiv\ln(e^Xe^Y)=X+\Big[\int_0^1\psi(e^{\text{ad}_X}e^{t\,\text{ad}_Y}))dt\Big]Y.
\label{BCH}
\ee
Here we have introduced the function $\psi(x)=\frac{x\ln x}{x-1}$ and alsothe  adjoint operator $\text{ad}_X$, which generates the commutator when applying to any other operators
\be
\text{ad}_X(Y\dots)=[X,Y\cdots].
\ee
Note that for any matrices $X$ and $Y$, 
\be
\Big[\sum_{ij}\dc_i X_{ij}c_j, \sum_{kl}\dc_k Y_{kl}c_l\Big]=\sum_{ij}\dc_i\Big([X,Y]\Big)_{ij}c_j
\ee
Since the right hand side of Eq. (\ref{BCH}) are all commutators between $X$ and $Y$, we find that
\be
\sum_{ij}\dc_i Z_{ij}c_j&=&
\sum_{ij}\dc_i\Big(X+\Big[\int_0^1\psi(e^{\text{ad}_X}e^{t\,\text{ad}_Y}))dt\Big]Y\Big)_{ij}c_j\nonumber\\
&=&\mathcal{X}+\Big[\int_0^1\psi(\exp[\text{ad}_{\mathcal{X}}]\exp[t\,\text{ad}_{\mathcal{Y}}])dt\Big]\mathcal{Y}\nonumber\\
&=&\ln(e^{\mathcal{X}}e^{\mathcal{Y}}).
\ee
Here we have defined
\be
\mathcal{X}=\sum_{ij}\dc_i X_{ij}c_j,\quad \mathcal{Y}=\sum_{ij}\dc_i Y_{ij}c_j.\nonumber
\ee
Making use of the above results, we find that
\be
&&\textrm{Tr}\Big[\exp\Big(\sum_{ij}\dc_i X_{ij}c_j\Big)\exp\Big(\sum_{kl}\dc_k Y_{kl}c_l\Big)\Big]\nonumber\\
&&=\textrm{Tr}\Big[\exp\Big(\sum_{ij}\dc_i Z_{ij}c_j\Big)\Big]
=\det(\mathbb{1}+e^Xe^Y)
\ee
This completes the proof of Eq. (\ref{eq-XY}).

\section{Details of EGP}\label{sec-EGPApp}
The matrix $A$ is an $N \times N$ matrix with non-zero blocks located on the sub-diagonal line as
\begin{equation}
A=\left(\begin{array}{cccccc}
   	0 & A_{2} & 0 & 0 & 0 & 0\\
   	0 & 0 & A_{3} & 0 & 0 & 0\\
   	0 & 0 & 0 & A_{4} & 0 & 0\\
   	0 & 0 & 0 & 0 & \ddots & 0\\
   	0 & 0 & 0 & 0 & 0 & A_{N}\\
   	A_{1} & 0 & 0 & 0 & 0 & 0
\end{array}\right).
\end{equation}
It is easy to verify that $A$ has the property $\textrm{Tr}A^n\neq0$ only for $n=mN$, where $m$ is an integer. The reason is that only the $N$th power of $A$ has nonzero diagonal blocks $A^N=(\prod_kA_k)\otimes\mathbb{1}$. With this property, we can simplify Eq. (\ref{egp4}) as
\begin{equation}
\begin{aligned}
\textrm{Tr}\Big[\sum_{n=1}^\infty (-1)^{n-1}\frac{A^n}{n}\Big]&
=\textrm{Tr}\Big[\sum_{m=1}^\infty (-1)^{mN-1}\frac{A^{mN}}{mN}\Big]\\
&=\textrm{tr}\Big[\sum_{m=1}^\infty(-1)^{mN-1}\frac{(\prod_kA_k)^m}{m}\Big]\\
&=\textrm{tr}\ln(1+M_T).\label{egp5}
\end{aligned}
\end{equation}
In the second line of above derivations, we have used the fact that
\be
&&\textrm{Tr}(A^N)^m=\textrm{Tr}[(\prod_kA_k)\mathbb{1}]^m
=N\textrm{tr}(\prod_kA_k)^m.
\ee
Here ``Tr'' means the trace over both momentum and orbital spaces while ``tr'' means trace over only orbital space.

\section{Verification of numerical accuracy}\label{sec-solvable}
\begin{figure}
\centering
\includegraphics[width=0.4\textwidth]{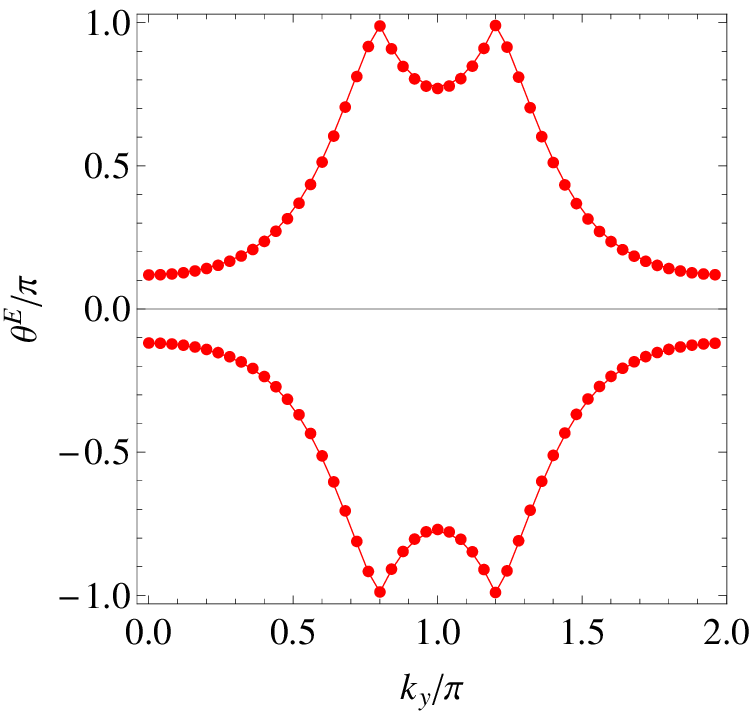}
\caption{Phase $\theta^E$ of the eigenvalues of $M_T(k_y)$ for the model of Eq. (\ref{sim}) as a function of $k_y$. The dots are numerical results while the line is the analytic result. Here $m=0.8$ and $T\Delta k=1$.}
\label{cmp}
\end{figure}

To verify the accuracy of the numerical results, we consider the following simple two-band Hamiltonian.
\be
H=\cos k_x\sigma_1+\sin k_x\sigma_2+(m+\cos k_y)\sigma_3.
\label{sim}
\ee
It is possible to give an analytical expression for the ensemble Wilson-loop. For fixed $k_y$, we consider an ensemble Wilson line, defined as
\be
V(k_x)=\mathcal{P}\exp\Big(-\frac{1}{T\Delta k}\int_0^{k_x} H(k'_x) d k'_x\Big).
\ee
To simplify the notation, we suppress the $k_y$ dependence of $V(k_x)$. The desired result is then given by $M_T=V(k_x=2\pi)$. Clearly, $V(k_x)$ can be solved from the following equation
\be
&&\frac{d V(k_x)}{d k_x}=-\frac{1}{T\Delta k} H(k_x)V(k_x)\nonumber\\
&&=-\frac{1}{T\Delta k}\Big[(m+\sin k_y)\sigma_3+e^{-ik_x\sigma_3/2}\sigma_1e^{i k_x\sigma_3/2}\Big]
V(k_x). \nonumber
\ee
One can simplify the above equation by a gauge transformation $V=e^{-i k_x\sigma_3/2}U$, and then the equation becomes
\be
\frac{d U(k_x)}{d k_x}=-\Big(\frac{1}{T\Delta k}\Big[(m+\sin k_y)\sigma_3+\sigma_1\Big]+\frac i2\sigma_3\Big)U(k_x). \nonumber
\ee
For fixed $k_y$, the term inside the big parenthesis is a constant matrix, it is straightforward to find that
\be
&&M_T=V(2\pi)\nonumber\\
&&=-\exp\Big(-\frac{2\pi}{T\Delta k}\Big[(m+\cos k_y)\sigma_3+\sigma_1\Big]-\pi i\sigma_3\Big).
\label{MT-2}
\ee
As a comparison, we plot the phases $\theta^E$ of the eigenvalues of $M_T(k_y)$ for the model of Eq. (\ref{sim}) as a function of $k_y$ in Figure \ref{cmp}.
The red dots are numerical results of Eq. (\ref{MT-1}) while the red line is computed from the analytical result of Eq. (\ref{MT-2}). One can see the two results agree with each other well. We remark that, however, this model is topologically trivial.

\bibliographystyle{apsrev}

\end{document}